\documentclass[a4paper]{revtex4}

\usepackage{dcolumn}
\usepackage{bm}
\usepackage{amsfonts}
\usepackage{graphicx,graphics}
\usepackage{amssymb}
\usepackage{rotate}
\usepackage{subfigure}
\usepackage{epsf}
\usepackage{epsfig}
\usepackage{amsmath}
\usepackage{dcolumn}
\def \t {tetrahedron }
\def \ts {tetrahedra }

\begin{document}

\title{\textbf{The Kasteleyn transition in three dimensions:  spin ice in a  $[100]$ field}}


\author{L. D. C. Jaubert$^{\dag}$, J.T. Chalker$^{\#}$, P. C. W. Holdsworth$^{\dag}$ and R. Moessner$^{\ast}$}

\address{$^{\dag}$Universit\'{e} de Lyon, Laboratoire de Physique, \'{E}cole Normale Sup\'{e}rieure de Lyon, 46 All\'{e}e d'Italie, 69364 Lyon cedex 07, France.}

\address{$^{\#}$Rudolf Peierls Centre for Theoretical Physics, 1
Keble Rd., Oxford, OX1 3NP, UK}

\address{$^{\ast}$Max-Plack-Institut f\"ur Physik komplexer Systeme, 01189 Dresden, Germany.}

\date{\today}

\begin{abstract}

We discuss the nearest neighbour spin ice model in the presence of a magnetic field placed along the cubic $[100]$ direction. As recently shown in {\it Phys. Rev. Lett. {\bf 100}, 067207, 2008}, the symmetry sustaining ordering transition observed at low temperature is a three dimensional Kasteleyn transition. We confirm this with numerical data using a non-local algorithm that conserves the topological constraints at low temperature and from analytic calculations from a Bethe lattice of corner sharing \ts . We present a thermodynamic description of the Kasteleyn transition and discuss the relevance of our results to recent neutron scattering experiments on spin ice materials.

\end{abstract}
\maketitle


\section{Introduction}

This transition was introduced by Kasteleyn~\cite{Kasteleyn63} in the context of ordering of hard core \textit{2D} dimers lying on the bonds of a honeycomb lattice. Applying local chemical potentials, $\mu_i$,  for dimers oriented in one of the $i=1,2,3$ possible directions leads to a singular ordering transition for finite values of the fugacities $z_i=exp(-\beta\mu_i)$. The transition can also occur for dimers on three dimensional equivalents of the honeycomb lattice such as the brick lattice \cite{Bhattacharjee83}, while the exponents can be determined in dimension $d$ through mapping onto a directed polymer problem \cite{Bhattacharjee91}. It has been extensively used to describe trans-gauche structural transitions in polymerized lipid bilayers and the resulting theory  provides a good qualitative description of the singular density change occurring in solvent bilayer systems as a function of temperature \cite{Nagle73}.  A Kasteleyn transition also occurs in the five vertex model on a square lattice \cite{Watson}, which maps onto the square ice model, satisfying the ice rules in two dimensions with direct relevance to spin ice materials\cite{Harris97}. It has been shown that the same Kasteleyn transition exists in the nearest neighbour model for spin ice materials on a pyrochlore lattice, with magnetic field directed close to the body centred cubic $[111]$ direction\cite{Moessner03}. Placing the field along this  direction isolates  layers orthogonal to the field, each of which  forms a kagom\'e lattice of directed tetrahedra whose vertices lie alternately above and below the plane.  The magnetic field fixes only one out of the four sublattices with the result that the extensive ground state degeneracy of zero field is only partially lifted. The ground state entropy remains finite and the kagom\'e spin problem maps exactly onto Kasteleyn's original dimer problem on the Honeycomb lattice. Tilting the field off the $[111]$ direction, giving a finite perpendicular field component $\vec H_{\perp}$, is equivalent to applying chemical potentials $\mu_i$ to the dimers and the spins order in the  kagom\'e planes for a finite value of  $\vec H_{\perp}/k_BT$ via Kasteleyn's transition. Ground state configurations of this model require the imposition of local constraints, which lead to power law dipole like correlations \cite{Isakov04,Henley05}.
Recent neutron studies of Ho$_2$Ti$_2$$_7$ with this field configuration \cite{Fennell07}  have shown experimentally the existence of these correlations in the form of well defined pinch points in reciprocal space for the intensity of the elastic diffuse scattering. In two dimensions these points should include logarithmically divergent intensities. In the presence of a perpendicular field component the singularities are predicted\cite{Moessner03} to drift with changing field strength, $\vec H_{\perp}$. It has recently been shown that the ordering transition seen when a field is placed along the $[100]$ direction \cite{Harris98} is a three dimensional Kasteleyn transition \cite{Jaubert08}. Magnetic measurements on Holmium and Dysprosium Titanate\cite{Fennell05,Fukazawa02} indicate that experimental systems could show strong features of the transition, but as we will see below, accessing it while remaining in thermodynamic equilibrium is a considerable experimental challenge.

\section{The Kasteleyn transition in spin ice in a [100] field}
\label{s:thermo}

In this paper we review the behaviour of the nearest neighbour spin ice model, with magnetic field placed along the cubic $[100]$ direction. Nearest neighbour moments (spins) of unit length are constrained to lie along the body centred crystal field directions of the tetrahedra: $\vec S_i = \pm \vec d_i$ and
interact with a ferromagnetic exchange interaction $J$ and with an external magnetic field, $\vec H$ in reduced units. The Hamiltonian~\cite{Harris97},  is
\begin{equation}\label{eq:hamiltonian}
\mathcal{H}= -\sum_{\langle ij \rangle} J \vec S_i . \vec S_j - \vec H .
\sum_i \vec S_i\,.
\end{equation}
The crystal fields frustrate the ferromagnetic exchange leading to a highly degenerate ground state satisfying the so called ice rules \cite{Bernal33} with two spins pointing in and two spins pointing out of each tetrahedron. 
This is the topological constraint which  takes the form $\vec \nabla . \vec S(\vec r)=0$ \cite{Isakov04} in a continuum description.
In zero field there is no phase transition down to zero temperature but,  in the presence of a magnetic field and in the limit where the topological constraints are rigorously applied the system orders at low temperature via a three dimensional \cite{Jaubert08} Kasteleyn transition.

The remarkable thing about the Kasteleyn transition is that it occurs when the system is rigorously constrained to the manifold of spin ice states such that the internal energy for each microstate in the reduced phase space has the same value, $U= -NJ/3$.  This allows for a particularly simple thermodynamic description which we present below. Just as for a paramagnet the magnetic Helmholtz free energy for the constrained system has an entropic component only: $F(N,T,M)=-TS(N,T,M)$, where $M$ is the magnetic order parameter. In a regular paramagnetic system $S$ goes to zero, as $M$ reaches its maximum value, $M_{max}$ but it does so with infinite slope and hence an infinite value for $H/k_BT$, where $H=|\vec H|$ is the field conjugate to $M$:
\begin{equation}
H=-\frac{1}{N}\frac{\partial F}{\partial M}.
\end{equation}
That is, there is no phase transition. However the hidden divergence free constraint can drive the entropy to zero for finite $H/k_BT$ giving a singular longitudinal susceptibility, $\chi=\partial M/\partial H$-this is the Kasteleyn transition. The Gibbs potential is defined $G^{\ast}=F-NMH$ from which  one can construct a Landau free energy, $G_L$ by expanding $G^{\ast}$ in small $x=M_{max}-M$ near the transition: 
%
\begin{equation}
\frac{G_L}{N}=(H-H_K)x + \frac{\alpha_2}{2} T x^2 + \frac{\alpha_3}{3} T x^3 +\dots -M_{max}h,
\end{equation}\label{Landau}
where $H_K$ is the critical field for $T$ fixed and where $\alpha_i$ are constants. If $\alpha_2 > 0$ one finds the mean field critical behaviour for fixed $H$:
\begin{equation}\label{betadef}
x=x_0(T-T_{\rm K})^{\beta},\;\;\; \beta=1,
\end{equation}
where $T_{\rm K}$ is the transition temperature for fixed $H$. One also finds that the specific heat, $C_{\rm H}$, and  susceptibility, $\chi$, jump discontinuously to zero at $T_{\rm K}$. For $T<T_{\rm K}$ $G^{\ast}(M)$ has no minimum for $0<M<M_{max}$.  $\partial G^{\ast} /\partial M$ is finite and negative at $M=M_{max}$, consistently with there being zero fluctuations and the entropy being strictly zero:  the system is perfectly ordered and represents a complete vacuum for excitations. Hence the transition is violently asymmetric, with fluctuations above the transition but not below. Through this asymmetry the transition has often been described as having both $1^{st}$ and $2^{nd}$ order characteristics\cite{Nagle89}. However from the above thermodynamics it is clearly second order, or continuous, within the Ehrenfest classification\cite{Stanley}.

\begin{figure}
     \centering
     \subfigure[]{
          \includegraphics[scale=0.8]{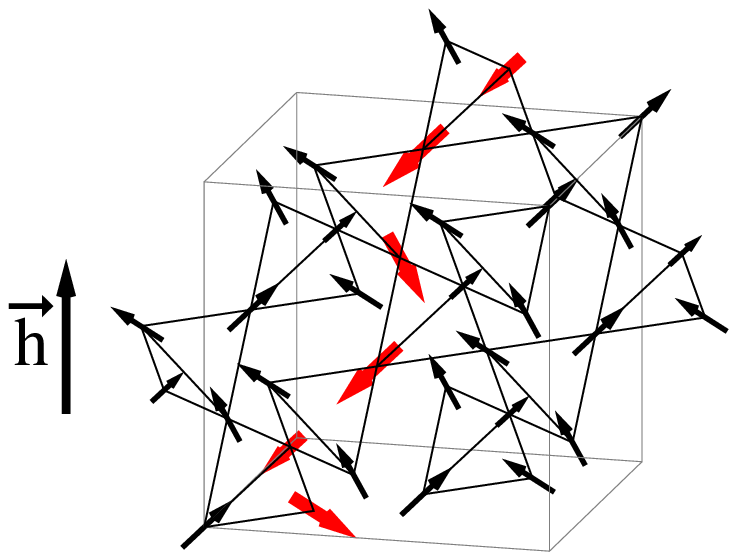}\label{fig:cell-1}} 
     \subfigure[]{
          \label{fig:Mh}
          \includegraphics[scale=0.5]{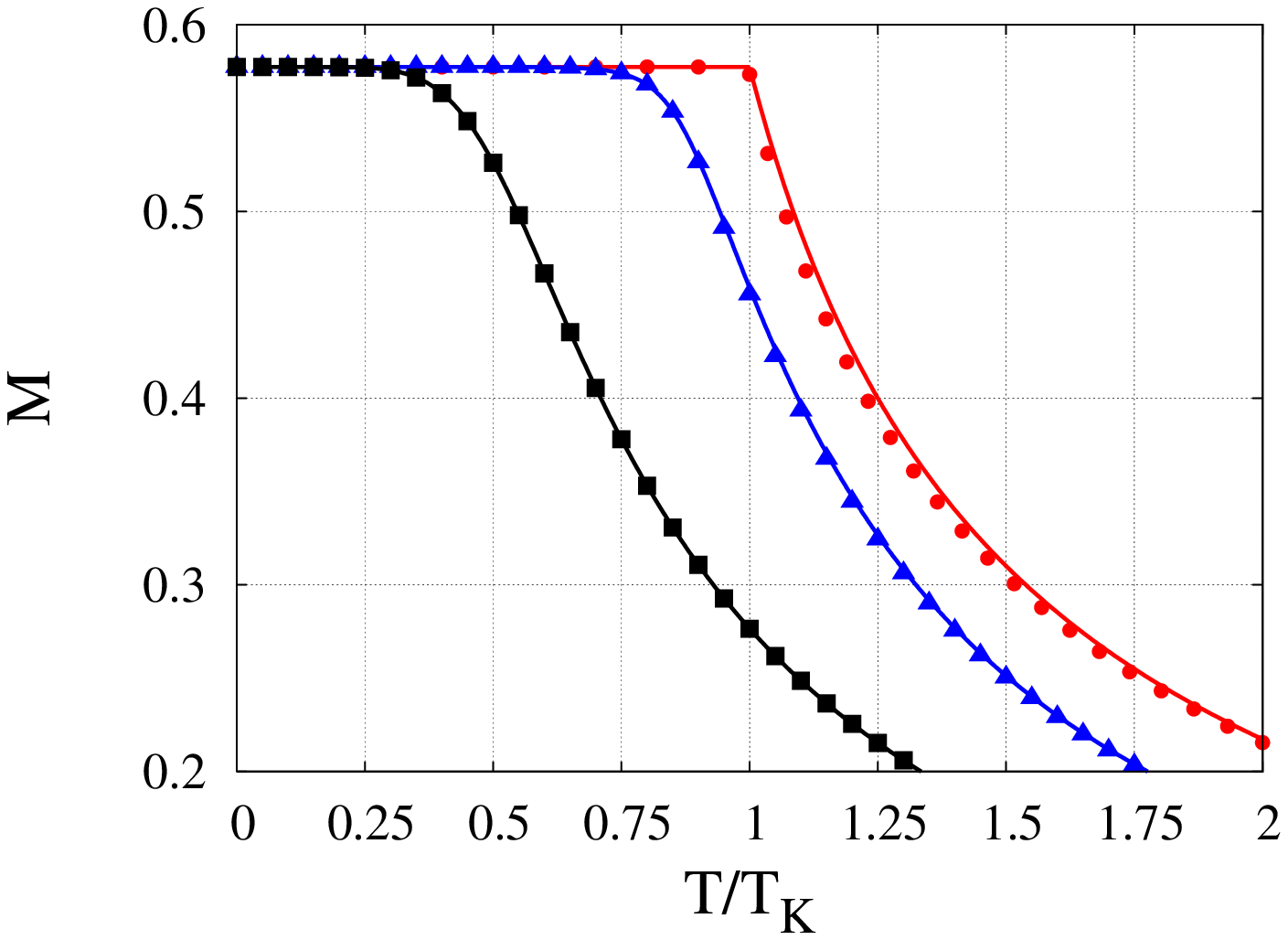}}
     
     \caption{$(a)$ The pyrochlore lattice with spins satisfying the ice rules. $\vec H$ is given along the $[100]$ direction. Black spins: long range ordered state, red spins: a string of flipped spins.\\
     \\
   {\bf Figure 1}  $(b)$(Color online) Magnetization per spin $M$ vs. $T/T_{\rm K}$ obtained from simulations for the pyrochlore lattice (dots) and analytically on the Bethe lattice (solid lines) for $H/J=10^{-3}$($\bullet$)$,\:0.13$($\blacktriangle$) and $0.58$($\blacksquare$).
     }
\end{figure}

Zero entropy can only be maintained at finite temperature if the topological constraints are rigorously applied ($J>>k_BT$) so that the lowest energy excitations off the perfectly ordered state are extended, with infinitely large energy in the thermodynamic limit. 
To see that this is the case consider a strong field at low temperatures which moves the system from a disordered spin ice state to the long range ordered state illustrated by the black spins in Fig \ref{fig:cell-1}. Relaxing back to a disordered state while maintaining the ice rules requires the flipping of a string of spins spanning the entire system (in the following discussion we impose periodic boundary conditions, in which case the string closes through the boundaries). To see this consider flipping a single spin out of the ordered state. This breaks the ice rules creating a pair of "topological defects" \cite{castelnovo08}, tetrahedra with three spins in and one out and vise versa, which can only be annealed by interacting with each other. Creating the defects costs magnetic enthalpy $4J/3+ 2H/\sqrt{3}$, but moving them further apart costs no further exchange energy in the nearest neighbour model. The exchange energy can be regained if the ensuing string passes right through the system and the topological defects destroy each other, returning the system to the constrained manifold of spin ice states. Such a string is illustrated by the red spins in Figure \ref{fig:cell-1}. Ignoring the exchange energy, each step of the string costs Zeeman energy $2H/\sqrt(3)$ but also represents an entropy gain of $\ln(2)$ as the string can leave the tetrahedron in one of two directions. If we define $L$ as the number of spins in the [100] direction, the total free enthalpy change for placing a string in the system is then
\begin{eqnarray}\label{eq:kasteleyn}
{\delta G}&=& L \left(\frac{2H}{\sqrt{3}}-T\ln2\right) \nonumber\\
&\propto& L\left(T_{\rm K}-T\right),\quad\textrm{where} \nonumber\\
T_{\rm K}&=&\frac{2H/\sqrt(3)}{\ln2} \label{eq:TK},
\end{eqnarray}
and strings appear above the temperature $T_{\rm K}$ only. The transition could be first order if strings attracted each other  but two strings passing though the same tetrahedron clearly have a reduced entropy over that for two separated strings. Hence, strings repel each other and the result is a continuous Kasteleyn transition at temperature $T_{\rm K}$. 

A consequence of the topological constraints and the broken magnetic symmetry is that a mapping exists between the strings of returned spins and world lines for bosons living in a $d-1$ dimensional space perpendicular to the field \cite{Jaubert08,Powell08}. The pyrochlore lattice has six fold cubic symmetry, which is reduced by the magnetic field, giving a unique definition for the arrow of time of the world lines along an imaginary time axis. The bosonic nature of the particles shows up in the passage of two strings through a single \t. The entropic contribution of the \t is zero and it is impossible to index strings "$1$" and "$2$" either side of the encounter without a fixed convention. However, with such a convention an exact mapping exists between spin and string configurations. In the absence of a field, there are six equivalent definitions for the arrow of time and a further convention is required to define the strings. 
In reference \cite{Jaubert08,Powell08} it was shown that through the definition of strings as world lines, the Kasteleyn transition in dimension $d$ maps onto a quantum phase transition in dimension $d-1$. The applied field corresponds to the boson chemical potential, $\mu \sim (h - h_K)$ and the transition is from a Bose condensed to a vacuum state at zero temperature. The upper critical dimension for this transition is $d=3$\cite{Fisher88}, which is compatible with previous analytic studies of the Kasteleyn transition in three dimensional dimer models\cite{Bhattacharjee83}. Hence, for the three dimensional Kasteleyn transition we expect near mean field behaviour, with  logarithmic corrections to scaling, while in two dimensions fluctuations should be important. This is indeed the case: the analytic solution of the two dimensional problem\cite{Moessner03} gives a value for the exponent, $\beta=1/2$.

\begin{figure}
     \centering
     \subfigure[]{
          \label{fig:ChM}
          \includegraphics[scale=0.5]{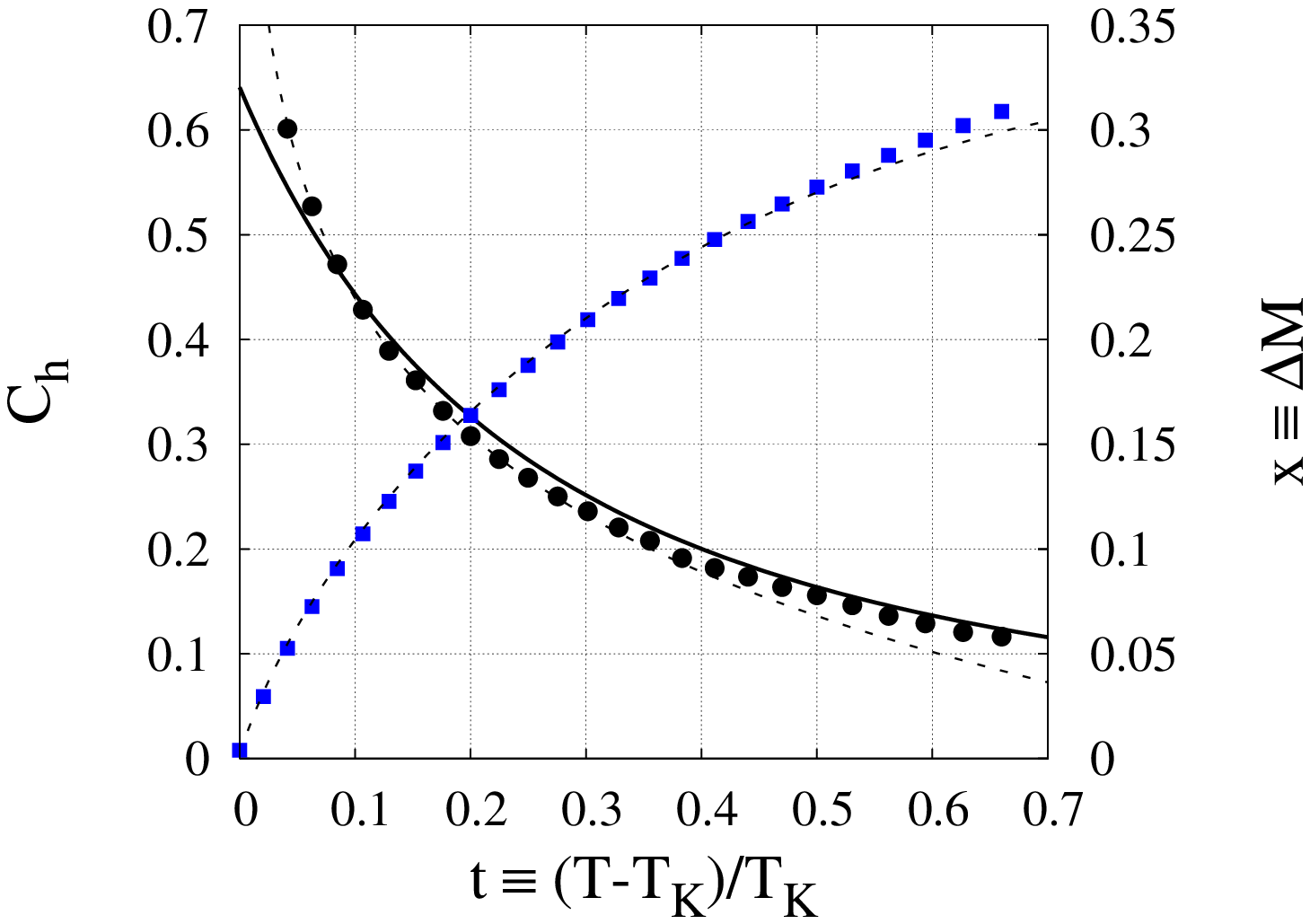}}
     \subfigure[]{
          \label{Correlation}
          \includegraphics[scale=0.5]{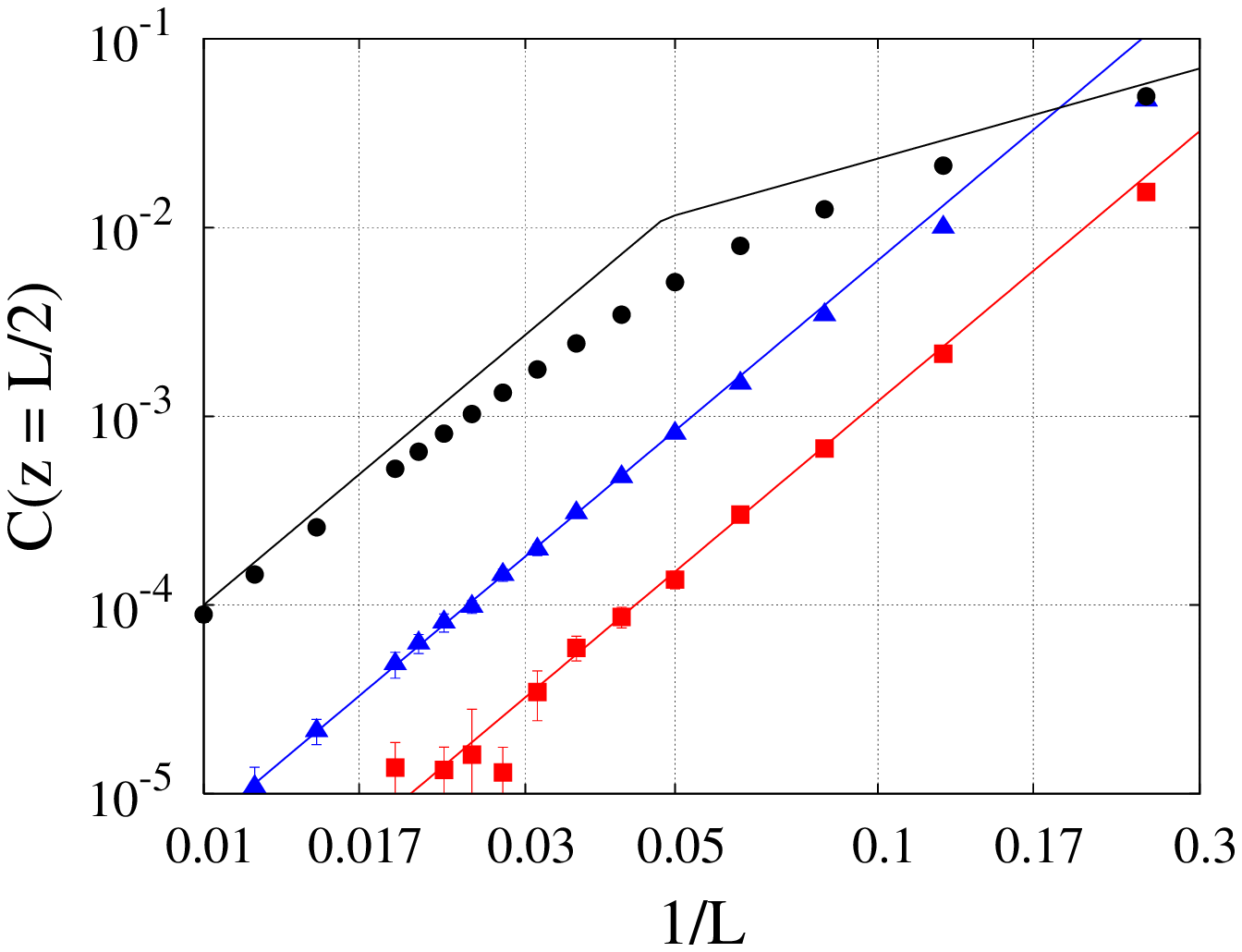}}
     
     \caption{$(a)$ (Color online) Specific heat $C_{\rm h}$ and relative magnetisation $\Delta M$ as a function of $t$ obtained from simulations ($\bullet$ for $C_{\rm h}$ and $\blacksquare$ for $\Delta M$) and analytically (solid line for $C_{\rm h}$) for $H = 10^{-3}  J$. The dashed lines are a fit to logarithmic singularities given in  eq.~(\ref{eq:ChM}).\\
     \\
   {\bf Figure 2}  $(b)$ (Color online) Finite-size scaling of spin correlations 
for $H/H_c=0.33$($\blacksquare$)$,\:0.66$($\blacktriangle$) and
  $0.93$($\bullet$), on log-log scale. Lines have slopes $3$ and $1$.    }
\end{figure}

We have tested these ideas using a non-local Monte Carlo algorithm which flips connected lines of spins \cite{Isakov04PRB,Jaubert08}. The lines  can form closed loops, either within the sample or through the boundaries, or they can transport topological defects non-locally. In the case of closed loops only, the algorithm restricts the system to fully constrained spin ice states, allowing an ergodic exploration of this manifold without passing through energetically unfavorable unconstrained states.
In Figure \ref{fig:Mh} we show $M = \frac{1}{N} |\sum_i \vec S_i |\; vs \; T$ for fixed  $H$. The points represent data generated by simulation using the non-local loop algorithm, while solid lines show analytic data from a calculation on a tree structure (Bethe lattice) of connected \t . The lack of closed loops here, allows for an essentially analytic solution of the problem \cite{Jaubert08b}. For $H/J=0.001$ the data show the characteristics of the Kasteleyn transition: asymmetric 
singular behaviour, with the magnetisation arriving at its saturation value $M_{max}=1/\sqrt{3}$ precisely  at $T=T_{\rm K}$. This is confirmed in figure \ref{fig:ChM}, where the magnetisation difference, $x=M_{max} - M$ and the  specific heat (and hence the susceptibility) are both seen to be singular.  As predicted through the mapping to the quantum phase transition, the simulation data for $C_{\rm H}$ fit accurately to a logarithmic divergence above the transition, while $x$ approaches zero linearly with $T-T_{\rm K}$ and has logarithmic corrections for larger
$x$:
\begin{equation}
x= \Delta M \propto t\left(1-a\ln(t)\right) \Leftrightarrow C_{\rm h}\propto - \ln(t) 
\label{eq:ChM}
\end{equation}
Detailed calculations for the Bethe lattice confirm the mean field picture developed phenomenologically above \cite{Jaubert08b}.

Previous numerical work found indications of a $1^{st}$ order phase transition at finite field~\cite{Harris98}. The observed first order nature is a  consequence of the loss of ergodicity due to the single spin flip dynamics used at that time. Interestingly experimental results~\cite{Fennell05} reflected the same out-of-equilibrium features, as natural dynamics is also local. However, with the introduction of the non-local algorithm the numerical results very clearly confirm that the transition is in fact a Kasteleyn transition similar to that observed for dimers on the brick lattice\cite{Bhattacharjee83}.

For larger fields the transition is rounded, this is because the topological  constraints are no longer rigorously imposed and the system can support a finite concentration of topological defects. Above a characteristic scale fixed by their separation the system will become regularly paramagnetic and this is just what is observed in the rounding of the transition.



Despite the fact that the constraints induce long range correlations the ferromagnetic susceptibility is not divergent away from the Kasteleyn transition. The coupling between spin and configurational space for dipolar interactions leads to both direction and sublattice dependence for the amplitude of the spin spin correlation function.  This gives a finite susceptibility at zero wave vector, $\vec Q=0$, but very strong variation at finite $\vec Q$, leading to the so called pinch points on the Brillouin zone boundaries \cite{Youngblood}. Although logarithmically divergent in 2D the intensity of the pinch points in three dimensions remains finite in zero field. 
Applying the appropriate magnetic field in the 2D problem breaks the square symmetry of zero field, introducing a length scale related to the symmetry breaking. As a result, the pinch points move in towards the Brillouin zone centre. Their arrival at the zone centre corresponds to the Kasteleyn transition. In 3D the pinch points remain fixed in reciprocal space but their intensity evolves with applied field, with the point at the Brillouin zone centre developing a logarithmic divergence at the transition.
This evolution can be seen directly in real space through spin-spin correlation functions $C(\vec r)= <\vec S_i(\vec r).\vec S_i(0)>$.  In Figure \ref{Correlation} we show finite size scaling plots of $C(z=L/2)$ for three dimensional systems of size $L$ for magnetic field along the $z$ axis, from which we can extract the power law behaviour of $C(\vec r)$ in the thermodynamic limit. For small field we find a $1/r^3$ behaviour consistently with the dipolar correlations leading to the pinch points. As one approaches the transition we observe a crossover to $1/r$ dependence. This is the regime where spin fluctuations along $z$ are due to multiple encounters with a single string. Here, mapping either to the quantum problem or  to a random walk problem lead to the $1/r$ behaviour observed in the figure\cite{Jaubert08,Powell08}.

%

\begin{figure}
     \centering
     \subfigure[]{
          \label{fig:fsize}
          \includegraphics[scale=0.5]{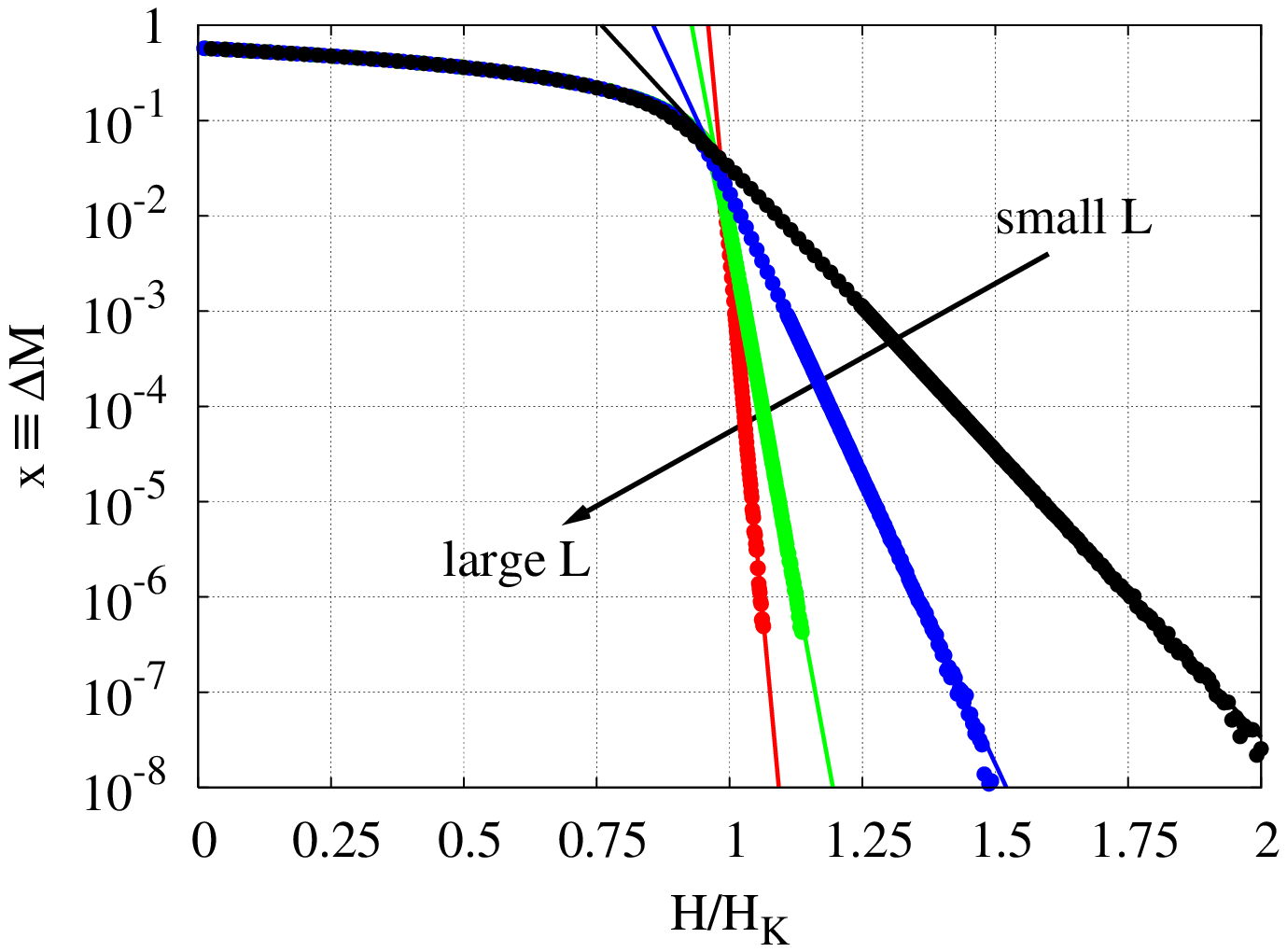}}
     \subfigure[]{
          \label{fig:Tom}
          \includegraphics[scale=0.5]{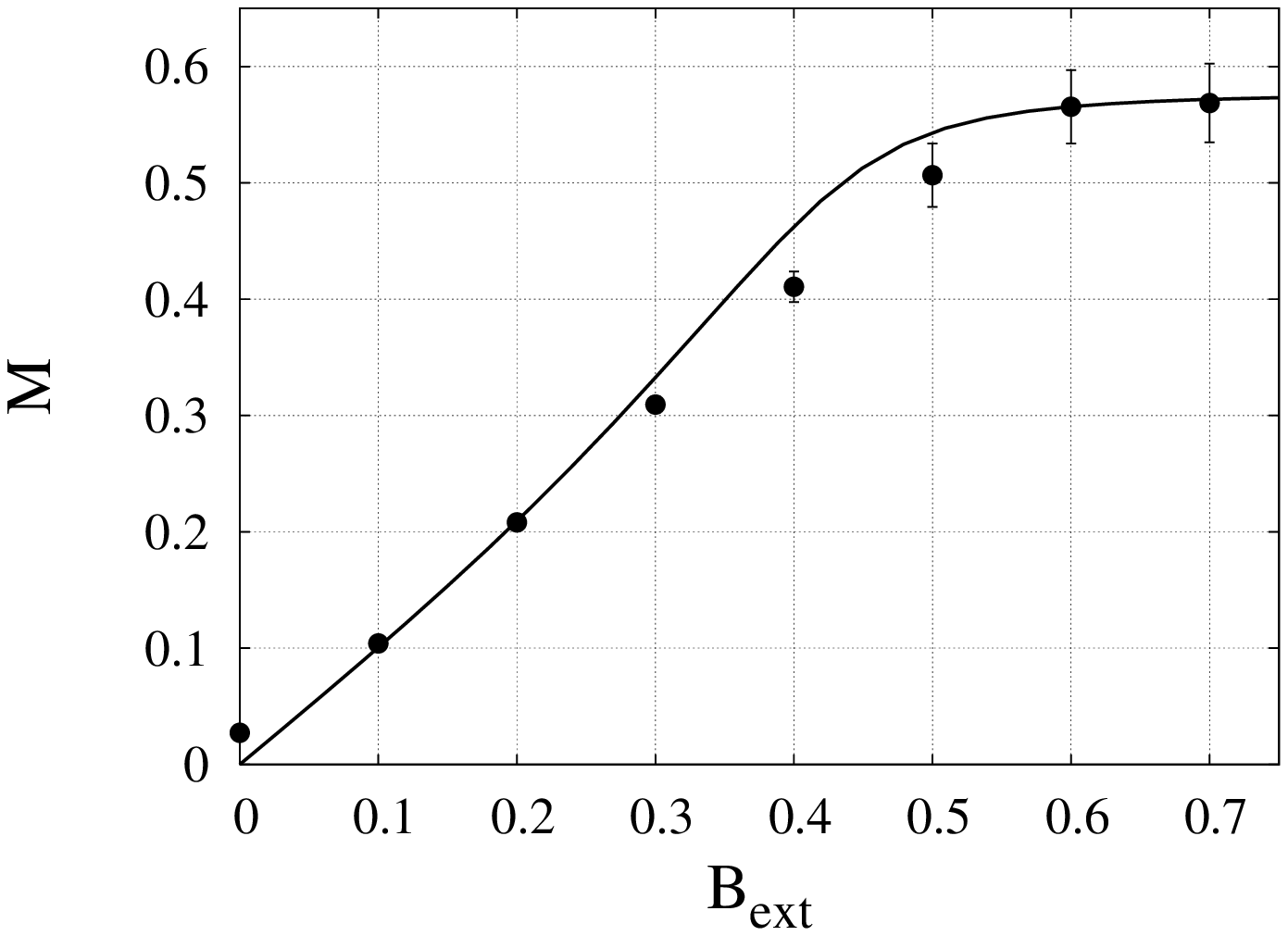}}
     
     \caption{ (Color online) Relative magnetisation $\Delta M$ as a function of $H/H_K$ obtained numerically ($\bullet$) for different sizes of the system: $L=20,40,100,200$. The straight lines are a fit to the Boltzmann factor $e^{\left(L\,\ln 2\right)\:H/H_K}$.\\
     \\
     {\bf Figure 3}  $(b)$ $M$ vs $B$ obtained analytically (solid lines) and
 experimentally (dots) by neutron scattering on the compound
 Ho$_2$Ti$_2$O$_7$ at $T=1.2K$. For fitting details, see main text.}
\end{figure}


The loop algorithm allows particularly accurate studies of finite size effects close to the transition where the density of strings is small. In the ordered regime, the free energy cost for the excitation of a string (\ref{eq:kasteleyn}), $ \delta G$ is dominated by the energy $E_{string}=2\,H\,L/\sqrt{3}$ hence the 
Boltzmann weight for string excitation decreases exponentially with system size: $w_{loop}\propto \exp\left(-\beta E_{string}\right)$ \cite{Isakov04PRB}. As the magnetisation difference $x$ is proportional to the number of strings one can observe this exponential finite size scaling by making an accurate measure of $x$ in the transition region. In Fig. (\ref{fig:fsize}) we show $x$ on a logarithmic scale, as a function of $T_{\rm K}/T$, for different system sizes. The data, averaged over up to $10^7$ configurations fit the exponential scaling remarkably well down to $x\sim 10^{-7}$. This analysis shows convincingly that the mechanism for the transition is indeed the generation of extensive excitations developed in this section.

\section{Discussion}

As the experimental systems fall out of equilibrium  at low temperature, it is difficult to compare theory with experiment very close to the Kasteleyn transition. But at reasonably low temperatures, where the concentration of topological defects is  low but sufficiently high to maintain ergodicity, it is possible to observe a rounded transition and to compare data with our theory. As an example we show, in Fig. \ref{fig:Tom}  one of the $[100]$ field scans performed at fixed low temperature
in Holmium Titanate~\cite{Fennell05}. The data is for  $T=1.2 {\rm K}$. This temperature is about $2/3$ of the effective nearest neighbour exchange constant $J_{eff} S^2/k_B \sim 1.8K$ estimated for Ho$_2$Ti$_2$O$_7$~\cite{Hertog00}. The magnetisation flattens off abruptly at the saturation value, $M_{exp}=10\mu_B/\sqrt{3}$ in a way reminiscent of the Kasteleyn transition but is indeed rounded near saturation. Also shown in the figure is our data calculated analytically from the Bethe lattice approximation for the pyrochlore system with the same ratio of $J/k_BT$. The scale of the magnetic field (that is changing from $|\vec H|$ in reduced units to $B$ in Tesla) was estimated after demagnetisation effects were taken into account on the  experimental data. A final scale factor of $1.6$ was required to make the comparison. The comparison is qualitatively good, suggesting that the data of Ref.~\cite{Fennell05} do provide evidence for a rounded
Kasteleyn transition in Ho$_2$Ti$_2$O$_7$, which could be more closely
approached in future experiments.



We have enjoyed many fruitful discussions with
 S.T. Bramwell and T. Fennell as well as with C. Castelnovo, 
M.J. Harris and R. Melko. PCWH acknowledges financial
support from the European Science Foundation PESC/RNP/HFM, from the
Royal Society and from the London Centre of
Nanotechnology. LJ acknowledges support from European Science
Foundation PESC/RNP/HFM and from ANR grant 05-BLAN-0105. Both thank
the Rudolph Peierls Centre for Theoretical Physics, University of
Oxford, where this work was begun.

\section*{References}

\bibliographystyle{apsrev}
\bibliography{bibli}

\begin{thebibliography}{24}
\expandafter\ifx\csname natexlab\endcsname\relax\def\natexlab#1{#1}\fi
\expandafter\ifx\csname bibnamefont\endcsname\relax
  \def\bibnamefont#1{#1}\fi
\expandafter\ifx\csname bibfnamefont\endcsname\relax
  \def\bibfnamefont#1{#1}\fi
\expandafter\ifx\csname citenamefont\endcsname\relax
  \def\citenamefont#1{#1}\fi
\expandafter\ifx\csname url\endcsname\relax
  \def\url#1{\texttt{#1}}\fi
\expandafter\ifx\csname urlprefix\endcsname\relax\def\urlprefix{URL }\fi
\providecommand{\bibinfo}[2]{#2}
\providecommand{\eprint}[2][]{\url{#2}}

\bibitem[{\citenamefont{Kasteleyn}(1963)}]{Kasteleyn63}
\bibinfo{author}{\bibfnamefont{P.~W.} \bibnamefont{Kasteleyn}},
  \bibinfo{journal}{J. Math. Phys. {\bf 4}, 287}  (\bibinfo{year}{1963}).

\bibitem[{\citenamefont{Bhattacharjee et~al.}(1983)\citenamefont{Bhattacharjee,
  Nagle, Huse, and Fisher}}]{Bhattacharjee83}
\bibinfo{author}{\bibfnamefont{S.~M.} \bibnamefont{Bhattacharjee}},
  \bibinfo{author}{\bibfnamefont{J.~F.} \bibnamefont{Nagle}},
  \bibinfo{author}{\bibfnamefont{D.~A.} \bibnamefont{Huse}}, \bibnamefont{and}
  \bibinfo{author}{\bibfnamefont{M.~E.} \bibnamefont{Fisher}},
  \bibinfo{journal}{J. Stat. Phys. {\bf 32}, 361}  (\bibinfo{year}{1983}).

\bibitem[{\citenamefont{Bhattacharjee and Rajasekaren}(1991)}]{Bhattacharjee91}
\bibinfo{author}{\bibfnamefont{S.~M.} \bibnamefont{Bhattacharjee}}
  \bibnamefont{and} \bibinfo{author}{\bibfnamefont{J.~J.}
  \bibnamefont{Rajasekaren}}, \bibinfo{journal}{Phys. Rev. A {\bf 44}, 6202}
  (\bibinfo{year}{1991}).

\bibitem[{\citenamefont{Nagle}(1973)}]{Nagle73}
\bibinfo{author}{\bibfnamefont{J.~F.} \bibnamefont{Nagle}},
  \bibinfo{journal}{Proc. Nat. Acad. Sci. USA {\bf 70}, 3443}
  (\bibinfo{year}{1973}).

\bibitem[{\citenamefont{Watson}(1999)}]{Watson}
\bibinfo{author}{\bibfnamefont{G.~I.} \bibnamefont{Watson}},
  \bibinfo{journal}{J. Stat. Phys. {\bf 94}, 1045}  (\bibinfo{year}{1999}).

\bibitem[{\citenamefont{Harris et~al.}(1997)\citenamefont{Harris, Bramwell,
  McMorrow, Zeiske, and Godfrey}}]{Harris97}
\bibinfo{author}{\bibfnamefont{M.~J.} \bibnamefont{Harris}},
  \bibinfo{author}{\bibfnamefont{S.~T.} \bibnamefont{Bramwell}},
  \bibinfo{author}{\bibfnamefont{D.~F.} \bibnamefont{McMorrow}},
  \bibinfo{author}{\bibfnamefont{T.}~\bibnamefont{Zeiske}}, \bibnamefont{and}
  \bibinfo{author}{\bibfnamefont{K.~W.} \bibnamefont{Godfrey}},
  \bibinfo{journal}{Phys. Rev. Lett. {\bf 79}, 2554}  (\bibinfo{year}{1997}).

\bibitem[{\citenamefont{Moessner and Sondhi}(2003)}]{Moessner03}
\bibinfo{author}{\bibfnamefont{R.}~\bibnamefont{Moessner}} \bibnamefont{and}
  \bibinfo{author}{\bibfnamefont{S.~L.} \bibnamefont{Sondhi}},
  \bibinfo{journal}{Phys. Rev. B {\bf 68}, 064411}  (\bibinfo{year}{2003}).

\bibitem[{\citenamefont{Isakov et~al.}(2004{\natexlab{a}})\citenamefont{Isakov,
  Gregor, Moessner, and Sondhi}}]{Isakov04}
\bibinfo{author}{\bibfnamefont{S.~V.} \bibnamefont{Isakov}},
  \bibinfo{author}{\bibfnamefont{K.}~\bibnamefont{Gregor}},
  \bibinfo{author}{\bibfnamefont{R.}~\bibnamefont{Moessner}}, \bibnamefont{and}
  \bibinfo{author}{\bibfnamefont{S.~L.} \bibnamefont{Sondhi}},
  \bibinfo{journal}{Phys. Rev. Lett. {\bf 93}, 167204}
  (\bibinfo{year}{2004}{\natexlab{a}}).

\bibitem[{\citenamefont{Henley}(2005)}]{Henley05}
\bibinfo{author}{\bibfnamefont{C.~L.} \bibnamefont{Henley}},
  \bibinfo{journal}{Phys. Rev. B {\bf 71}, 014424}  (\bibinfo{year}{2005}).

\bibitem[{\citenamefont{Fennell et~al.}(2007)\citenamefont{Fennell, Bramwell,
  McMorrow, Manuel, and Wildes}}]{Fennell07}
\bibinfo{author}{\bibfnamefont{T.}~\bibnamefont{Fennell}},
  \bibinfo{author}{\bibfnamefont{S.~T.} \bibnamefont{Bramwell}},
  \bibinfo{author}{\bibfnamefont{D.~F.} \bibnamefont{McMorrow}},
  \bibinfo{author}{\bibfnamefont{P.}~\bibnamefont{Manuel}}, \bibnamefont{and}
  \bibinfo{author}{\bibfnamefont{A.}~\bibnamefont{Wildes}},
  \bibinfo{journal}{Nature Physics {\bf 3}, 566}  (\bibinfo{year}{2007}).

\bibitem[{\citenamefont{Harris et~al.}(1998)\citenamefont{Harris, Bramwell,
  Holdsworth, and Champion}}]{Harris98}
\bibinfo{author}{\bibfnamefont{M.~J.} \bibnamefont{Harris}},
  \bibinfo{author}{\bibfnamefont{S.~T.} \bibnamefont{Bramwell}},
  \bibinfo{author}{\bibfnamefont{P.~C.~W.} \bibnamefont{Holdsworth}},
  \bibnamefont{and} \bibinfo{author}{\bibfnamefont{J.~D.~M.}
  \bibnamefont{Champion}}, \bibinfo{journal}{Phys. Rev. Lett. {\bf 81}, 4496}
  (\bibinfo{year}{1998}).

\bibitem[{\citenamefont{Jaubert
  et~al.}(2008{\natexlab{a}})\citenamefont{Jaubert, Chalker, Holdsworth, and
  Moessner}}]{Jaubert08}
\bibinfo{author}{\bibfnamefont{L.}~\bibnamefont{Jaubert}},
  \bibinfo{author}{\bibfnamefont{J.~T.} \bibnamefont{Chalker}},
  \bibinfo{author}{\bibfnamefont{P.~C.~W.} \bibnamefont{Holdsworth}},
  \bibnamefont{and} \bibinfo{author}{\bibfnamefont{R.}~\bibnamefont{Moessner}},
  \bibinfo{journal}{Phys. Rev. Lett {\bf 100}, 067207}
  (\bibinfo{year}{2008}{\natexlab{a}}).

\bibitem[{\citenamefont{Fennell et~al.}(2005)\citenamefont{Fennell, Pentrenko,
  Fak, Garnder, Bramwell, and Ouladdiat}}]{Fennell05}
\bibinfo{author}{\bibfnamefont{T.}~\bibnamefont{Fennell}},
  \bibinfo{author}{\bibfnamefont{O.~A.} \bibnamefont{Pentrenko}},
  \bibinfo{author}{\bibfnamefont{B.}~\bibnamefont{Fak}},
  \bibinfo{author}{\bibfnamefont{J.~S.} \bibnamefont{Garnder}},
  \bibinfo{author}{\bibfnamefont{S.~T.} \bibnamefont{Bramwell}},
  \bibnamefont{and}
  \bibinfo{author}{\bibfnamefont{B.}~\bibnamefont{Ouladdiat}},
  \bibinfo{journal}{Phys. Rev. B {\bf 72}, 224411}  (\bibinfo{year}{2005}).

\bibitem[{\citenamefont{Fukazawa et~al.}(2002)\citenamefont{Fukazawa, Melko,
  Higashinaka, Maeno, and Gingras}}]{Fukazawa02}
\bibinfo{author}{\bibfnamefont{H.}~\bibnamefont{Fukazawa}},
  \bibinfo{author}{\bibfnamefont{R.~G.} \bibnamefont{Melko}},
  \bibinfo{author}{\bibfnamefont{R.}~\bibnamefont{Higashinaka}},
  \bibinfo{author}{\bibfnamefont{Y.}~\bibnamefont{Maeno}}, \bibnamefont{and}
  \bibinfo{author}{\bibfnamefont{M.~P.~J.} \bibnamefont{Gingras}},
  \bibinfo{journal}{Phys. Rev. B {\bf 65}, 054410}  (\bibinfo{year}{2002}).

\bibitem[{\citenamefont{Bernal and Fowler}(1933)}]{Bernal33}
\bibinfo{author}{\bibfnamefont{J.~D.} \bibnamefont{Bernal}} \bibnamefont{and}
  \bibinfo{author}{\bibfnamefont{R.~H.} \bibnamefont{Fowler}},
  \bibinfo{journal}{J. Chem. Phys. {\bf 1}, 515}  (\bibinfo{year}{1933}).

\bibitem[{\citenamefont{Nagle et~al.}(1989)\citenamefont{Nagle, Yokoi, and
  Bhattacharjee}}]{Nagle89}
\bibinfo{author}{\bibfnamefont{J.~F.} \bibnamefont{Nagle}},
  \bibinfo{author}{\bibfnamefont{C.~S.~O.} \bibnamefont{Yokoi}},
  \bibnamefont{and} \bibinfo{author}{\bibfnamefont{S.~M.}
  \bibnamefont{Bhattacharjee}}, \emph{\bibinfo{title}{Phase Transitions and
  Critical Phenomena, Ed. C. Domb and J. L. Lebowitz}}
  (\bibinfo{publisher}{Academic Press}, \bibinfo{year}{1989}),
  chap.~\bibinfo{chapter}{2}.

\bibitem[{\citenamefont{Stanley}(1971)}]{Stanley}
\bibinfo{author}{\bibfnamefont{H.~E.} \bibnamefont{Stanley}},
  \emph{\bibinfo{title}{Introduction to Phase transitions and Critical
  Phenomena}} (\bibinfo{publisher}{OUP}, \bibinfo{year}{1971}).

\bibitem[{\citenamefont{Castelnovo et~al.}(2008)\citenamefont{Castelnovo,
  Moessner, and Sondhi}}]{castelnovo08}
\bibinfo{author}{\bibfnamefont{C.}~\bibnamefont{Castelnovo}},
  \bibinfo{author}{\bibfnamefont{R.}~\bibnamefont{Moessner}}, \bibnamefont{and}
  \bibinfo{author}{\bibfnamefont{S.~L.} \bibnamefont{Sondhi}},
  \bibinfo{journal}{Nature {\bf 451}, 42}  (\bibinfo{year}{2008}).

\bibitem[{\citenamefont{Powell and Chalker}(2008)}]{Powell08}
\bibinfo{author}{\bibfnamefont{S.}~\bibnamefont{Powell}} \bibnamefont{and}
  \bibinfo{author}{\bibfnamefont{J.~T.} \bibnamefont{Chalker}},
  \bibinfo{journal}{cond-mat/0803.4204}  (\bibinfo{year}{2008}).

\bibitem[{\citenamefont{Fisher and Hohenberg}(1988)}]{Fisher88}
\bibinfo{author}{\bibfnamefont{D.~S.} \bibnamefont{Fisher}} \bibnamefont{and}
  \bibinfo{author}{\bibfnamefont{P.~C.} \bibnamefont{Hohenberg}},
  \bibinfo{journal}{Phys. Rev. B {\bf 37}, 4936}  (\bibinfo{year}{1988}).

\bibitem[{\citenamefont{Isakov et~al.}(2004{\natexlab{b}})\citenamefont{Isakov,
  Raman, Moessner, and Sondhi}}]{Isakov04PRB}
\bibinfo{author}{\bibfnamefont{S.~V.} \bibnamefont{Isakov}},
  \bibinfo{author}{\bibfnamefont{K.~S.} \bibnamefont{Raman}},
  \bibinfo{author}{\bibfnamefont{R.}~\bibnamefont{Moessner}}, \bibnamefont{and}
  \bibinfo{author}{\bibfnamefont{S.~L.} \bibnamefont{Sondhi}},
  \bibinfo{journal}{Phys. Rev. B {\bf 70}, 104418}
  (\bibinfo{year}{2004}{\natexlab{b}}).

\bibitem[{\citenamefont{Jaubert
  et~al.}(2008{\natexlab{b}})\citenamefont{Jaubert, Holdsworth, and
  Moessner}}]{Jaubert08b}
\bibinfo{author}{\bibfnamefont{L.}~\bibnamefont{Jaubert}},
  \bibinfo{author}{\bibfnamefont{P.~C.~W.} \bibnamefont{Holdsworth}},
  \bibnamefont{and} \bibinfo{author}{\bibfnamefont{R.}~\bibnamefont{Moessner}},
  \bibinfo{journal}{Unpublished}  (\bibinfo{year}{2008}{\natexlab{b}}).

\bibitem[{\citenamefont{Youngblood and Axe}(1981)}]{Youngblood}
\bibinfo{author}{\bibfnamefont{R.~W.} \bibnamefont{Youngblood}}
  \bibnamefont{and} \bibinfo{author}{\bibfnamefont{J.~D.} \bibnamefont{Axe}},
  \bibinfo{journal}{Phys. Rev. B {\bf 23}, 232}  (\bibinfo{year}{1981}).

\bibitem[{\citenamefont{den Hertog and Gingras}(2000)}]{Hertog00}
\bibinfo{author}{\bibfnamefont{B.~C.} \bibnamefont{den Hertog}}
  \bibnamefont{and} \bibinfo{author}{\bibfnamefont{M.~J.~P.}
  \bibnamefont{Gingras}}, \bibinfo{journal}{Phys. Rev. Lett. {\bf 84}, 3430}
  (\bibinfo{year}{2000}).

\end{thebibliography}

\end{document}